\documentclass[multphys,vecphys]{svmult}

\usepackage{makeidx}         
\usepackage{graphicx}        
\usepackage{multicol}        
\usepackage{cite}            
\usepackage[bottom]{footmisc}

\makeindex             

\usepackage{amsmath,amssymb}

\newcommand{\sref}[1]{Section~\ref{#1}}

\newcommand{\fref}[1]{Fig.~\ref{#1}}

\newcommand{\figcite}[1]{{\protect \cite{#1}}}

\newcommand{\normfig}{0.7\textwidth}
\newcommand{\largefig}{0.85\textwidth}
\newcommand{\datafig}{0.55\textwidth}

\newcommand{\vf}{v_F}
\newcommand{\kf}{k_F}

\begin{document}

\title{From Luttinger to Fermi liquids in organic conductors}
\author{T. Giamarchi}
\institute{DPMC-MaNEP, University of Geneva, 24 Quai Ernest
Ansermet, 1211 Geneva, Switzerland \\
\texttt{Thierry.Giamarchi@physics.unige.ch}}

\maketitle

This chapter reviews the effects of interactions in quasi-one
dimensional systems, such as the Bechgaard and Fabre salts, and in
particular the Luttinger liquid physics. It discusses in details how
transport measurements both d.c. and a.c. allow to probe such a
physics. It also examine the dimensional crossover and deconfinement
transition occurring between the one dimensional case and the higher
dimensional one resulting from the hopping of electrons between
chains in the quasi-one dimensional structure.

\section{Introduction}

Organic conductors, such as TMTTF and TMTSF compounds offer unique
challenges. Indeed from the theoretical point of view most of our
understanding of interacting electronic problems is based on
Landau's Fermi liquid (FL) theory
\cite{landau_fermiliquid_theory_static,nozieres_book,abrikosov_book}.
However it is well known that the effects of interactions can be
greatly enhanced by reduced dimensionality. In one dimension,
interactions destroy the Fermi liquid and lead to a quite different
state known as a Luttinger liquid (LL) \cite{giamarchi_book_1d}. For
commensurate systems such as the organic conductors, interactions
can also lead to a Mott insulating (MI) state, another state showing
clearly the effects of strong correlations.

The organic conductors are thus natural candidates to search for the
existence and properties of such states. However because  of their
very three dimensional nature, they provide not a single one
dimensional electron gas, but a very large number of such one
dimensional systems coupled together. This allows for a unique new
physics to emerge where the system is able to crossover from a one
dimensional behavior to a more conventional three dimensional one
\cite{giamarchi_review_chemrev}. This richness is also a drawback,
since it is now an important issue to know whether MI or LL physics
can be realized at all in these systems. Although for the members of
the TMTTF family it was soon undisputable that they are indeed Mott
insulators and that interaction effects were important
\cite{bourbonnais_book_organics}, the situation was far from being
clear for the TMTSF compounds, that were behaving as good metals
with many characteristics of a nice Fermi liquid. Despite important
efforts no convincing experimental case could be made for LL
behavior and the nature of the normal phase and of the crossover
scale between a one dimensional and higher dimensional behavior of
these compounds remained hotly debated. Understanding the nature of
the normal phase and the effect of interactions in these compounds,
in addition of being important in its own right, was of course
potentially crucial in connection with the low temperature ordered
phase and in particular of the superconducting one.

Theoretical progress in computing the transport properties
\cite{giamarchi_umklapp_1d,giamarchi_mott_shortrev} and
corresponding experimental measurements of the optical conductivity
\cite{dressel_optical_tmtsf,schwartz_electrodynamics} allowed for a
solution of this dilemma and proved the LL properties of the above
mentioned organic conductors. In addition, this led to a definite
reexamination
\cite{giamarchi_mott_shortrev,giamarchi_review_chemrev} of what was
commonly believed  as the main reason
\cite{emery_umklapp_dimerization} for the Mott insulating nature of
the parent compounds, namely the dimerization of the organic chain.
It also allowed to clearly determine the crossover scale between the
one and higher dimensional behavior in these systems, leading to a
very consistent understanding of the physics of these materials as
well as the one of quarter filled compounds
\cite{kato_quarter_synthesis,heuze_quarterfilled_refs,bourbonnais_book_organics}.

I will thus focus in this chapter on the issue of the transport in
quasi-one dimensional organic conductors and how it can be used to
probe for the MI and LL physics, and more generally on the question
of dimensional crossover and deconfinement between the low
dimensional Luttinger liquid or Mott insulator and a more
conventional high dimensional metal. I mostly concentrate here to
the specific applications to the organics here and refer the reader
to the review of C. Bourbonnais and D. Jerome
\cite{bourbonnais_book_organics} for a general introduction,
specific experimental data and references on the quasi-one
dimensional organics and to previous literature for more details on
the derivations
\cite{giamarchi_umklapp_1d,giamarchi_mott_shortrev,giamarchi_book_1d},
further theoretical issues
\cite{giamarchi_mott_shortrev,giamarchi_review_chemrev} and
references.

The plan of this chapter is as follows. In \sref{sec:basicint} I
review the basic questions and concepts for a system of coupled one
dimensional chains. In \sref{sec:luttingermott} I discuss the
transport properties of isolated chains and how one can use them to
probe for MI and LL physics, as well as various characteristics of
the interactions in these systems. In \sref{sec:coupledchains} I
discuss effects specifically due to the coupling between the chains,
such as the deconfinement transition and some transverse transport
properties. Finally conclusions and perspectives are presented in
\sref{sec:conclusions}.

\section{General ideas} \label{sec:basicint}

Let me first summarize the main ideas and challenges in connection
with the quasi-one dimensional nature of the organics, and the
observation of LL behavior.

\begin{figure}
 \centerline{\includegraphics[width=\largefig]{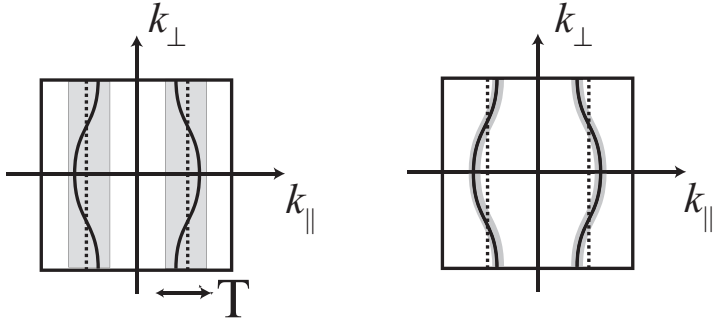}}
 \caption{Dimensional crossover for noninteracting electrons. $k_\parallel$ is the momentum along
 the chains and $k_\perp$ the one perpendicular to the chains. (left) If the
 temperature $T$ (or any other external energy scale, represented by the gray area)
 is larger than the warping of the Fermi surface due to interchain hopping the system cannot feel the
 warping. It is thus behaving as a one-dimensional system. (right)
 At a lower temperature/energy the system feels the
 two (or three) dimensional nature of the dispersion and thus behaves as a full two (or three) dimensional
 system. There is thus a dimensional crossover as the
 temperature/energy is lowered.}
 \label{fig:openfermi}
\end{figure}

\begin{figure}
 \centerline{\includegraphics[width=\largefig]{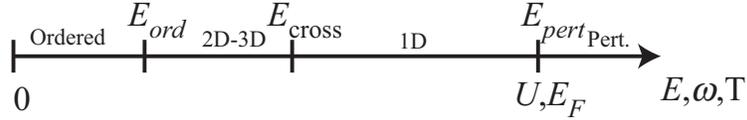}}
 \caption{Separation of energy scales if the interchain hopping
 $t_\perp$ is much smaller than the intrachain one $t_\parallel$.
 For energies larger than the intrachain hopping (or equivalently the Fermi energy $E_F$)
 and interactions (denoted generically $U$), simple perturbation
 theory is valid. Below this scale the system is in an interacting one dimensional regime. The interchain hopping
 couples the chains at an energy $E_{\rm cross}$ and destroys the one-dimensional physics.
 For non-interacting particles $E_{\rm cross} \sim t_\perp$ but this scale is renormalized by interactions
 into $t^\nu_\perp$ in a LL. For commensurate systems (Mott
 insulators) the Mott gap can suppress the single particle hopping
 and drive $E_{\rm cross}$ to zero. In all cases the system can
 have a transition to an ordered state at an energy $E_{\rm ord}$. If the dimensional crossover takes place before (as is the
 case shown in the above figure), this transition should be described from the two- or three- dimensional
 interacting theory.}
 \label{fig:separ}
\end{figure}

The chains are characterized by intra ($t_\parallel$) and interchain
($t_\perp$) single particle hopping. The main effect of the
interchain single particle hopping is to induce a dimensional
crossover between a one dimensional situation and a higher
dimensional one. In the absence of interactions such dimensional
crossover is easy to understand. In Fourier space the kinetic energy
becomes
\begin{equation} \label{eq:disper}
 \epsilon(k_\parallel,k_\perp) = - 2 t_\parallel \;\cos(k_\parallel a)
 - 2 t_\perp \;\cos(k_\perp b)
\end{equation}
where $b$ denotes a perpendicular direction. If the perpendicular
hopping $t_\perp$ is much smaller than the parallel one
$t_\parallel$, which is the relevant case for the quasi-one
dimensional organics, then (\ref{eq:disper}) leads to the open Fermi
surface of \fref{fig:openfermi}.
If one is at an energy scale (for example the temperature $T$ or the
frequency $\omega$) larger than the warping of the Fermi surface
then the warping is washed out, which means that no coherent hopping
can take place between the chains. In that case the system is
indistinguishable from one with a flat Fermi surface and can thus be
considered as a one-dimensional system. On the other hand if the
temperature or energy is much smaller than the warping of the Fermi
surface all correlation functions are sensitive to the presence of
the warping, and the system is two- or three-dimensional. Since I
considered free electrons in the above example, this crossover
occurs at an energy scale of the order of the interchain hopping, as
is summarized in \fref{fig:openfermi}. In presence of interactions
and commensurability, the problem is of course much more complicated
and interactions affect drastically this behavior compared to the
non interacting case. Indeed, for the organics $t_\perp$ is much
smaller than the intrachain characteristic energy scales such as the
kinetic energy or the interactions. In that case the chains can
experience the full effect of the interaction in a one dimensional
regime before the processes due to interchain hopping can spoil this
pure one-dimensional physics. The dimensional crossover scale must
thus be computed from the one dimensional interacting theory. This
is summarized in \fref{fig:separ}.

\begin{figure}
 \centerline{\includegraphics[width=\largefig]{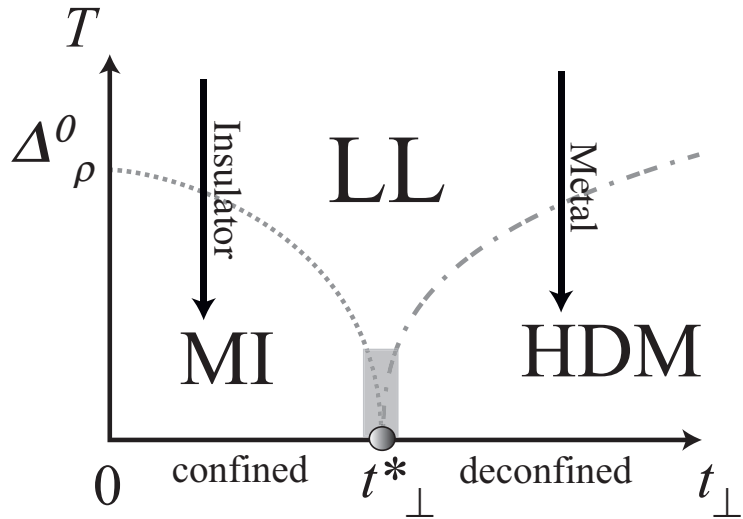}}
 \caption{Schematic phase diagram expected for coupled one
 dimensional commensurate chains as a function of the temperature $T$ (or an energy $E$) and the
 interchain hopping $t_\perp$. In the absence of $t_\perp$ the system is a one dimensional
 Mott insulator with a Mott gap $\Delta_\rho^0$. If $t_\perp$ is weak the ground state
 is still a Mott insulator (MI).
 For temperature larger than the Mott gap
 $\Delta(t_\perp)$ (dotted line) one observes a crossover to a Luttinger liquid (LL)
 regime. Beyond a critical value $t_\perp^*$ the system has a
 deconfinement transition towards a high dimensional
 metal (HDM). Additional complications can occur near this point as
 schematically represented by the gray box (see text).
 Above a certain crossover scale $T^*$ (dash-dotted line) the LL behavior
 is recovered since no coherent hopping can take place between
 chains. These two crossovers can be observed in
 different materials or upon application of pressure by lowering the temperature as indicated by
 the two arrows. The two confined and deconfined regions correspond
 respectively to apparent insulating and metallic behavior when the
 temperature is lowered.
 After \cite{biermann_dmft1d_hubbard_short,giamarchi_review_chemrev}.}
 \label{fig:deconf_phasediag}
\end{figure}

For the case of a commensurate system, the situation is even more
complicated and a phase diagram as a function of the temperature (or
another energy scale probing the system) and interchain hopping
$t_\perp$ is shown in \fref{fig:deconf_phasediag}
If the chains are uncoupled, $t_\perp = 0$, one recovers the
behavior of one dimensional interacting electrons. For an
incommensurate system this gives the well known Luttinger liquid
behavior, that I will discuss in more details in
\sref{sec:luttingermott}. The organics are commensurate systems, and
thus at low temperature one can expect to have a Mott insulator
behavior. An important question, that I will discuss further in
\sref{sec:luttingermott} concerns the origin of this Mott insulating
behavior. Indeed the material has a quarter filled band (of holes)
\cite{bourbonnais_book_organics}, but a small dimerization of the
order of $\Delta_d \sim 100K$ makes the band effectively half
filled. One has thus to determine which commensurability is
important for the Mott behavior. Regardless of this point, one must
get an insulating behavior for an isolated chain at low temperature,
characterized by a Mott gap in the charge excitation spectrum
$\Delta_\rho$. For temperatures $T \ll \Delta_\rho$ one thus sees
the Mott insulating behavior. This is the region represented as (MI)
in the phase diagram of \fref{fig:deconf_phasediag}. For
temperatures larger than the Mott gap $T \gg \Delta_\rho$ the Mott
behavior is not observable and one has a crossover to the Luttinger
liquid behavior in the isolated chain. This is denoted as (LL) in
the phase diagram of \fref{fig:deconf_phasediag}.

The interchain coupling brings the system from the one dimensional
behavior to a higher dimensional one. However, if each chain
develops a gap, it means that the single-particle Green's function
decays \emph{exponentially}. The single-particle hopping is now an
irrelevant variable. The formation of a gap is thus in direct
competition with the interchain hopping. For small interchain
hopping the system thus remains an insulator but with a smaller gap
$\Delta_\rho(t_\perp)$ than for an isolated chain. As far as single
particle hopping is concerned the system is thus essentially one
dimensional. This is the regime corresponding to the insulating part
of \fref{fig:deconf_phasediag}. As shown in \fref{fig:separ}, the
system can still in this regime undergo a transition to an ordered
state if particle-hole (density-density) or particle-particle
(Josephson coupling) interactions are present between the chains.
Such interactions are in any case generated to second order in the
single particle hopping.

The competition between the Mott physics and the interchain hopping
means that by increasing the interchain hopping to a critical value
one can break the one-dimensional Mott gap. Thus at $T=0$ a quantum
phase transition occurs for $t_\perp = t_\perp^*$  above which the
insulating Mott state is destroyed and the system becomes a high
dimensional metal. This transition is known as a deconfinement
transition where both the nature of the state as well as the
effective dimensionality of the system change. The properties and
consequences of this transition will be discussed in
\sref{sec:coupledchains}. Beyond this critical value of $t_\perp$
the low temperature properties are the ones of the high dimensional
metal (HDM). What is the nature of such a high dimensional metal
(and in particular whether it is a Fermi liquid or not) and how it
is affected by the fact that it stemmed from coupled one dimensional
chains is of course an important question. As the temperature is
increased above a temperature $T^*(t_\perp)$ one can expect a
crossover towards a one dimensional LL behavior again. Indeed at
high temperature coherent hopping between the chains cannot take
place. This dimensional crossover energy scale is drastically
affected by both the interactions present in the chains \emph{and}
the commensurability and is thus quite different from the one for
noninteracting electrons.

This generic phase diagram is directly relevant for the Q1D organics
\cite{bourbonnais_book_organics} due to their commensurate nature.
In particular they are very good realization of quasi-one
dimensional systems with hopping integrals of the order of $t_a
\simeq 3000K$, $t_b \simeq 300K$, $t_c \simeq 20K$, leading to
relatively well separated energy scales in which one is indeed
dominated by the intrachain hopping.  The band is quarter filled,
with a small dimerization along the chains giving some half filled
character to the system as well. The various parameters
($t_\parallel$, $t_\perp$ and the dimerization) can be tuned either
by changing the chemistry of the compound or by applying external
pressure, so the phase diagram of \fref{fig:deconf_phasediag} can be
roughly seen as a temperature (or energy) - pressure phase diagram.
I will come back to the role of pressure in
\sref{sec:luttingermott}. Experimentally, at ambient pressure, the
(TMTTF)$_2$PF$_6$ compound displays an insulating behavior (MI). A
transition to a metallic phase is found, with increasing pressure
and the properties of the TMTTF compounds evolve toward those of the
compounds of the TMTSF family, which are good conductors. This
evolution is clear from the a-axis resistivity measurements (see
e.g. Fig.~1.5 of \cite{bourbonnais_book_organics}). Such an
insulating behavior is well consistent with what one would expect
for a one-dimensional Mott insulator. The minimum of the resistivity
(followed by an activated law as temperature is lowered) defines the
onset of the MI regime on \fref{fig:deconf_phasediag}. It is thus
clear that the interactions play a crucial role in the TMTTF family
even at relatively high energies. For the TMTSF, the question is
more subtle in view of the metallic behavior at ambient pressure and
it was even suggested that such compounds could be described by a FL
behavior with weak interactions \cite{gorkov_sdw_tmtsf}. On the
contrary, interpretations of deviations of $1/T_1$ in NMR
\cite{bourbonnais_rmn} or magnetoresistance
\cite{behnia_tmtsfclo4_rmn} as due to a one-dimensional behavior
would suggest that one dimensional effects would persist to
temperatures as low as $20K$, a much too low scale compared to the
naive one given by the bare interchain hopping $t_b \sim 300K$. The
TMTTF and TMTSF family thus prompts for very fundamental questions
in connection with one dimensional physics:
\begin{enumerate}
 \item Are interactions also important for the metallic members of
 the family or can they be simply regarded as a Fermi liquid
 with an anisotropic Fermi surface?

 \item If indeed one can identify Luttinger liquid behavior,
 what are the Luttinger parameters?

 \item If the system is a Mott insulator is this mostly due to the
 dimerization of the band or is the quarter filling commensurability
 sufficient?

 \item What are the deconfinement scale $t_\perp^*$ and beyond that
 point the crossover scale $T^*$ below which the system is not one
 dimensional any more ? What is the nature of the high dimensional
 metallic phase ?
\end{enumerate}

I will now show how transport measurement, and specially optical
conductivity, have proven to be a key tool in addressing and to a
large extent answering these important questions.

\section{Mott insulators and 1D transport}
\label{sec:luttingermott}

Let me first examine the properties of the system in a regime where
the coherent hopping between the chains can be neglected. This is
the regime corresponding to the LL and MI parts of
\fref{fig:deconf_phasediag}. In that regime the properties are
essentially the ones of isolated chains, and in particular the
transport properties along the chains can be computed from a pure
one dimensional limit.

\subsection{Theory of transport}

I here recall only the salient points on transport in connection
with the quasi-one dimensional organics and refer the readers to
\cite{giamarchi_umklapp_1d,giamarchi_mott_shortrev,giamarchi_book_1d,giamarchi_review_chemrev}
for more details on the transport in one dimension and references.

If the filling is not commensurate, all excitations of a one
dimensional system are sound waves of density and spin density. A
convenient basis to describe such a system is provided by the
so-called bosonization technique \cite{giamarchi_book_1d}. The
energy of these excitations is given by a standard elastic-like
Hamiltonian. The Hamiltonian of the system is the sum of a part
containing only charge excitations and one containing only spin
excitations.
\begin{equation} \label{eq:hambas}
 H = H_\rho + H_\sigma
\end{equation}
where $H_{\nu}$ ($\nu=\rho,\sigma$) is of the form
\begin{equation}\label{eq:hamphi}
 H = \frac1{2\pi}\int dx \; [u_\nu K_\nu(\pi \Pi_\nu(x))^2 + \frac{u_\nu}{K_\nu}(\nabla\phi_\nu(x))^2]
\end{equation}
$\phi_\nu$ and $\Pi_\nu$ are conjugate variables
$[\phi_\nu(x),\Pi_\nu(x')] = i \hbar \delta(x-x')$. The fields
$\phi_\nu$ are related to the long wavelength distortions of the
charge $\rho(x)$ and spin $\sigma(x)$ electron density by
$\rho(x),\sigma(x) = - \sqrt2\nabla\phi_{\rho,\sigma}(x)/\pi$.
$u_{\rho,\sigma}$ are the velocities of these collective
excitations. In the absence of interactions $u_\rho = u_\sigma =
\vf$. Interactions of course renormalize the velocities of charge
and spin excitations, as in higher dimensions. $K_{\rho,\sigma}$ are
dimensionless parameters depending on the interactions. For systems
with spin rotation symmetry $K_\sigma = 1$ (for repulsive
interactions) while the spin excitations are gapped for attractive
interactions. $K_\rho = 1$ in the absence of interactions and quite
generally $K_\rho < 1$ for repulsive ones. The three parameters
$u_\rho$, $u_\sigma$ and $K_\rho$ completely characterize the low
energy properties of a one dimensional system. They can be computed
for a given microscopic model as a function of the interactions
\cite{luther_chaine_xxz,schulz_hubbard_exact,ogata_tj,mila_hubbard_etendu,kawakami_hubbard,kawakami_bethe_U<0},
but as was shown by Haldane
\cite{haldane_bosonisation_spin,haldane_bosonisation,Haldane_Luttinger},
the form (\ref{eq:hambas}) is the generic low energy form. This
means that (\ref{eq:hambas}) and the parameters $u_\rho$,
$u_\sigma$, $K_\rho$ play a role similar to the one of the Landau
Fermi liquid Hamiltonian (and Landau parameters) in higher
dimensions. To have again in one dimension a concept equivalent to
the Fermi liquid, i.e. a generic description of the low energy
physics (for energies lower than $E_{\rm pert}$ of \fref{fig:separ})
of the interacting problem, is of course extremely useful. This
removes part of the caricatural aspects of any modelization of a
true experimental system and allows to easily deal with extensions
such as the commensurability with the lattice.

The form (\ref{eq:hambas}) immediately shows that an excitation that
is looking like a free electron (i.e. that carries both charge and
spin) cannot exist. This is a very important difference between a
Luttinger and a Fermi liquid since in the latter, in addition to
collective modes of charge and spin, individual excitations
(quasiparticles) carrying both charge and spin and looking
essentially like a free electron do exist
\cite{landau_fermiliquid_theory_static,nozieres_book,abrikosov_book}.
In addition, the correlation functions in a LL display non universal
power laws with exponents dependant on the interactions via the
Luttinger parameter $K_\rho$. For example the single particle
correlation function decays with distance or time with an exponent
$\zeta = \frac14[K_\rho + K_\rho^{-1}] + \frac12$. The fact that it
decays faster than $1/r$ which is the case for free electrons or a
Fermi liquid shows directly that single particle excitations do not
exist in one dimension. Similarly spin-spin or density-density
correlations have a $2\kf$ oscillating part decaying with an
exponent $K_\rho+1$. Probing such power laws is thus a direct proof
of the LL behavior.

For commensurate systems one has to modify the Hamiltonian
(\ref{eq:hambas}) to take the commensurability with the lattice into
account. Such commensurability is at the root of the Mott
transition. Although one can of course work out the Mott transition
from microscopic models such as the Hubbard model
\cite{lieb_hubbard_exact}, the Luttinger liquid theory provides an
excellent framework to take into account the effects of a lattice
and describe the Mott transition. It is particularly well adapted
for the case of the organics since, as we will see, the Mott gap is
smaller than $E_{\rm pert}$ and thus the LL theory is indeed a
suitable starting point at these energies. To incorporate the Mott
transition in the Luttinger liquid description one must take into
account that in presence of a lattice the wavevector is in fact
defined modulo a vector of the reciprocal lattice (that is, in one
dimension a multiple of $2\pi/a$ with $a$ the lattice spacing).
Thus, in addition to the interaction processes that truly conserve
momentum $k_1+k_2 = k_3 + k_4$ one can now have umklapp processes
\cite{dzyaloshinskii_umklapp} such that $k_1+k_2 - k_3 - k_4 = Q$
where $Q$ is a vector of the reciprocal lattice. Since umklapps do
not conserve momentum they are the only ones that can lead to a
finite resistivity, and are responsible for the $T^2$ law in a Fermi
liquid \cite{ziman_phonon_book}. The umklapp process is also
responsible for the Mott transition in one dimension. It order for
such process to be efficient at the Fermi level it is necessary to
have $4\kf = 2\pi/a$ namely $\kf = \pi/2$ or one electron per site
(half filling). This corresponds to the case where two electrons are
scattered from one side of the Fermi surface ($-\kf$) to the other
side ($+\kf$). This is indeed the most standard case for having a
Mott insulator. But in fact, umklapps are not restricted to one
particle per site \cite{giamarchi_curvature,schulz_mott_revue}, but
occur for any commensurate fillings. Indeed, if $2p\kf = 2\pi q/a$
(where $p$ and $q$ are integers) then one can show that an
additional term must be added to (\ref{eq:hambas}). For even
commensurabilities ($p = 2n$), that corresponds to case of the
quasi-one dimensional organics) this term is
\cite{giamarchi_umklapp_1d,giamarchi_curvature,giamarchi_mott_shortrev}
\begin{equation} \label{eq:complet}
 H_{\frac1{2n}} = g_{\frac1{2n}} \int dx\; \cos(n \sqrt8 \phi_\rho(x))
\end{equation}
where $n$ is the order of the commensurability ($n=1$ for
half-filling --- one particle per site; $n=2$ for quarter-filling
--- one particle every two sites and so on). The coupling constant
$g_{1/2n}$ is the umklapp process corresponding to the
commensurability $n$.

\begin{figure}
 \centerline{\includegraphics[width=\largefig]{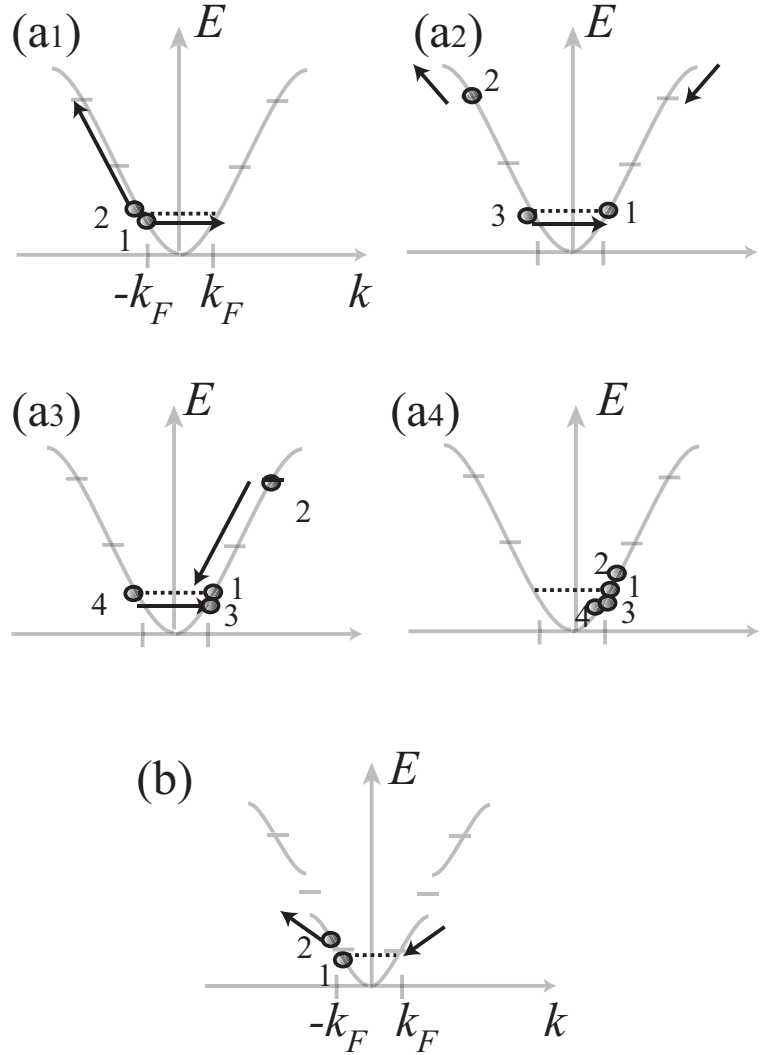}}
 \caption{Umklapp processes important for the organic conductors.
 (a 1-4)
 A quarter-filled umklapp can be constructed from a third-order perturbation theory
 in the interaction $U$. It consists in transferring four particles from one side of the Fermi surface to the other.
 The sequence of scattering due to the interaction is shown in
 figures (1-4). Three electrons are transferred from $-\kf$ to
 $+\kf$ by three successive interaction processes, while the
 momentum difference is absorbed by a fourth electron until it
 reaches the opposite side of the Fermi surface.
 The processus needs two intermediate states of high-energy $W$ of
 the order of the bandwidth. Thus for small interactions the amplitude for such a process is of order $U (U/W)^2$.
 (b)
 Dimerization opens a gap in the band. Because of this gap
 the quarter-filled band becomes effectively half filled. This reduces the zone boundary. The
 dimerization gap $\Delta_d$ thus creates even for a quarter-filled
 system an half filling umklapp where two particles can be transferred from one side of the Fermi surface to the other.
 The amplitude of such a process is proportional to the dimerization
 gap and thus of the order of $U \Delta_d/W$.
 (After \cite{giamarchi_review_chemrev}.)
 }
 \label{fig:umklapphigh}
\end{figure}

If the bosonization representation can give the universal form of
the Hamiltonian and the umklapp term, the amplitude of the umklapp
coefficients $g_{\frac1{2n}}$ depends on the precise microscopic
interaction. At half filling, for a Hubbard model, $g_{\frac1{2}}$
is of the order of the interaction $U$. Higher commensurability
umklapps can be estimated perturbatively. For a quarter-filled band
such that $8\kf = 2\pi/a$ (this corresponds to $n=2$ in the above
notations), to produce an umklapp one needs to transfer \emph{four}
particles from one side of the Fermi surface to the other to get the
proper $8\kf$ momentum transfer. This can be done in higher-order
perturbation terms by doing three scatterings as shown in
\fref{fig:umklapphigh}.
For weak interactions the amplitude of such a process would thus be
of order $U (U/W)^2$, where $W$ is the bandwidth. In addition to the
above process, there is an additional one for the Bechgaard or Fabre
salt family. Indeed in these systems the stack is slightly dimerised
\cite{jerome_review_chemrev,bourbonnais_book_organics}. This
dimerization opens a gap in the middle of the band as indicated in
\fref{fig:umklapphigh}. Thus although the system is originally
quarter filled the dimerization turns the system into a half-filled
band. This means that even if the system is quarter filled, a non
zero $g_{1/2}$ exists in addition to $g_{1/4}$. If $\Delta_d$ is the
dimerization gap the strength of such umklapp is $g_{1/2}^d = U
(\Delta_d/W)$. Note that contrarily to what happens in a true half
filled system the umklapp coefficient is now much smaller than the
typical interaction $U$. This allows to get a small Mott gap even if
the interactions are large. In presence of dimerization a quarter
filled system can thus be a Mott insulator either because of the
half-filling umklapp (that exists now because of the dimerization)
or because of the quarter-filled one. Which process is dominant
depends of course on the strength of the dimerization and of the
interactions, and has important consequences on the physics of the
system \cite{giamarchi_mott_shortrev}.

From (\ref{eq:complet}) all the properties of the Mott transition
and transport in a one dimensional system can be worked out. The
system is a Mott insulator for $K_\rho < K^*_\rho = 1/n^2$ where $n$
is the order of the commensurability. The larger the
commensurability the smaller $K_\rho$ needs to be for the system to
become insulating. For a commensurability $n=1$, that is,
half-filling the critical value is $K_\rho = 1$. This means that,
contrarily to the higher dimensional case, {\it any} repulsive
interactions turn the system into an insulator. For a quarter-filled
band ($n=2$) the critical value is $K_\rho = 1/4$. To get the
insulator one needs both pretty strong interactions \emph{and}
interactions of a finite range, since the minimum value of $K_\rho$
for a local interaction is $K_\rho=1/2$
\cite{schulz_conductivite_1d}. This is physically obvious: in order
to stabilize a structure in which there is a particle every two
sites one cannot do it with purely local interactions. The range of
the interactions in addition of their strength and thus the precise
chemistry of the compound controls the range of values of $K_\rho$
that one is able to explore.

\begin{figure}
   \centerline{\includegraphics[width=\normfig]{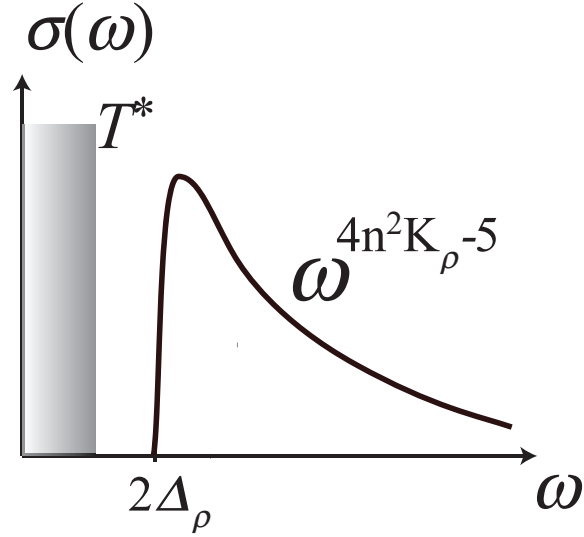}}
\caption{A.c. conductivity along the chains for a commensurability
of order $n$. $\Delta_\rho$ is the Mott gap. The full line is the
conductivity in the Mott insulator (the confined region). Above the
optical gap (twice the thermodynamic one $\Delta_\rho$) the
conductivity decays as a power law with an exponent $\mu = 4 n^2
K_\rho - 5$ characteristic of the Luttinger liquid behavior. A
simple band insulator would give $\omega^{-3}$. In the deconfined
region most of the features remain, except that below the
dimensional crossover scale $T^*$ the conductivity is not given by
the one dimensional theory any more. The metallic nature corresponds
to the appearance of a Drude peak close to zero frequency. This
Drude peak must be computed from a two- (or three-) dimensional
theory.}
\label{sigacdel0}
\end{figure}

\begin{figure}
   \centerline{\includegraphics[width=\largefig]{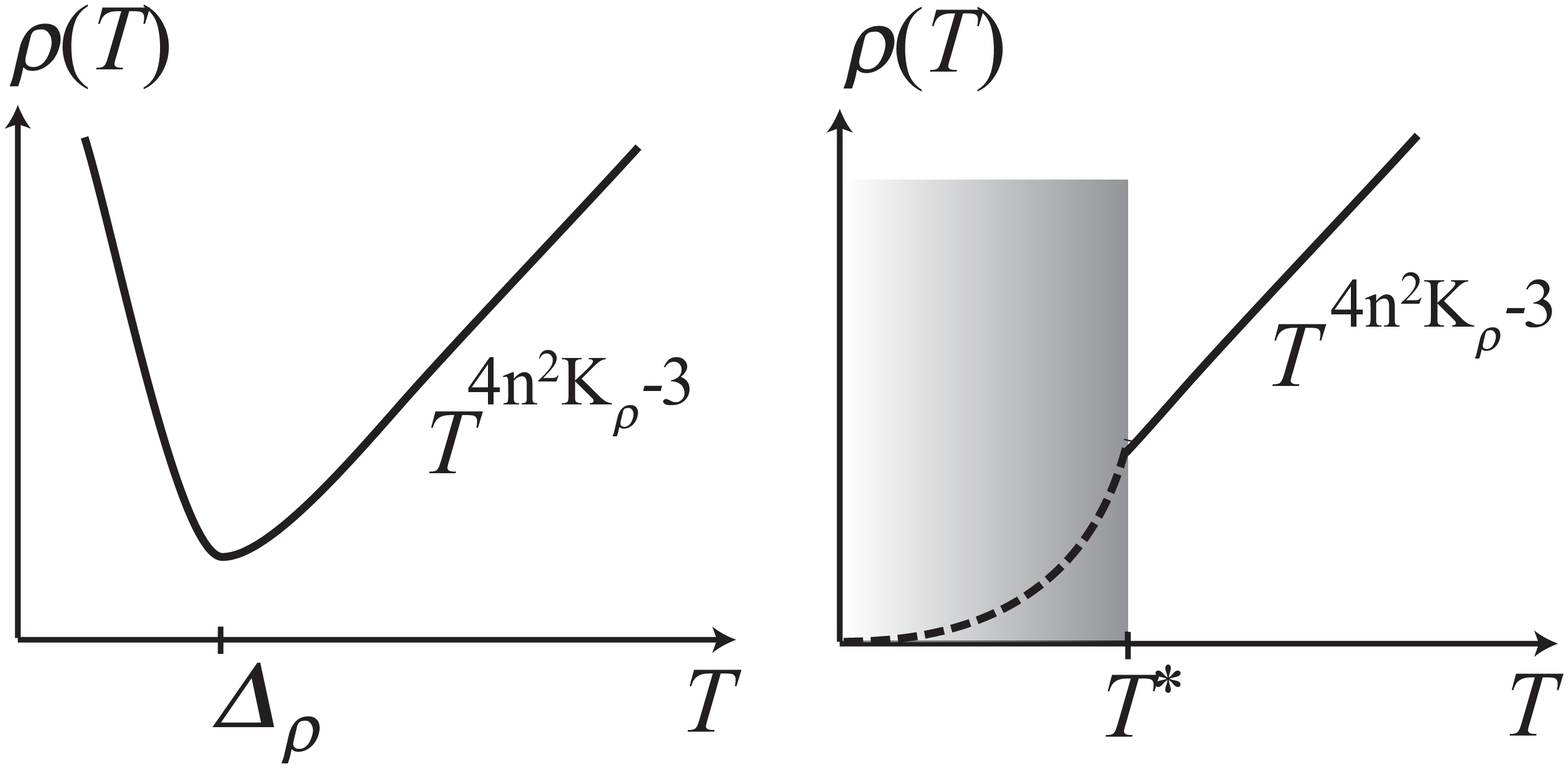}}
\caption{D.c. conductivity along the chains as a function of the
temperature $T$ for a commensurability of order $n$. $\Delta_\rho$
is the Mott gap. Left: confined region. Above the Mott gap the dc
transport shows an exponent $4n^2 K_\rho -3$ characteristic of the
Luttinger liquid. Below the Mott gap the number of carriers is
exponentially small, and any scattering will give an exponentially
small conductivity. Right: deconfined region. The Mott gap scale
does not exist any more. Above the dimensional crossover scale $T^*$
the temperature dependence is essentially identical to the one on
the left and shows LL behavior. Below the scale $T^*$ the system
must be described by a two- or three dimensional theory and one can
expect a temperature dependence much more conventional (it would be
$T^2$ for a simple Fermi liquid).} \label{dcsig}
\end{figure}

Of course the Mott transition and the Luttinger physics have drastic
consequences on the transport properties and one can expect quite
different properties than for Fermi liquid. Thus transport can be
used as an efficient probe. As we will see it allows to probe both
the single particle behavior (or absence thereof) and the Luttinger
liquid collective excitations \cite{giamarchi_umklapp_1d,giamarchi_mott_shortrev,%
giamarchi_attract_1d,giamarchi_curvature,%
controzzi_cond1d_form,jeckelmann_dmrg_conductivity_hubbard,rosch_conservation_1d}.
A schematic plot of the ac conductivity (at $T=0$) is shown in
\fref{sigacdel0}. In the Mott insulator $\sigma$ is zero until
$\omega$ is larger than the optical gap $2\Delta_\rho$. For
frequencies larger than the Mott gap, interactions dress the
umklapps and give a nonuniversal (i.e. interaction-dependent) power
law-like decay. Such a power law can be described by renormalization
group calculations  coupled to a memory function formalism
\cite{giamarchi_umklapp_1d,giamarchi_mott_shortrev}. The results of
this approach have been subsequently confirmed by form factor
calculations \cite{controzzi_cond1d_form}. If one ignores the
renormalization of $K_\rho$ by the umklapp (for the effect of the
renormalization of $K_\rho$ see \cite{giamarchi_umklapp_1d}) one
gets for the a.c. conductivity, for frequencies larger than the Mott
gap
\begin{equation} \label{eq:accond}
 \sigma(\omega) \sim \omega^{4 n^2 K_\rho -5}
\end{equation}
where $n$ is the order of commensurability.
The dc conductivity can be computed by the same methods
\cite{giamarchi_umklapp_1d,giamarchi_mott_shortrev,rosch_conservation_1d}
and is shown in \fref{dcsig}.
Here again the dressing of umklapps by the other interactions
results in a nonuniversal power law dependence for temperatures
larger than the Mott gap $\Delta_\rho$. Within the same
approximations than for (\ref{eq:accond}) one obtains
\begin{equation} \label{eq:dccond}
 \rho(T) \sim T^{4 n^2 K_\rho - 3}
\end{equation}

\subsection{Tranport in the organics}

\begin{figure}
 \centerline{\includegraphics[width=\normfig]{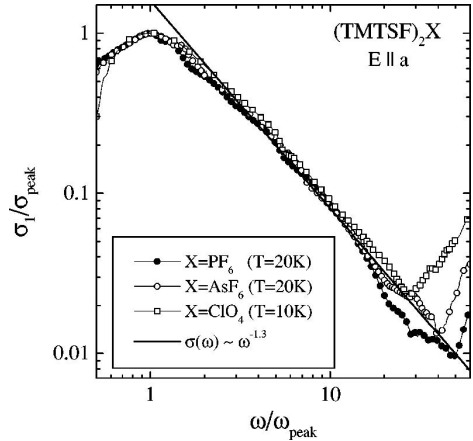}}
 \caption{Optical conductivity along the chain axis in the TMTSF family.
 The conductivity is rescaled by the gap in various samples. A
 power law behavior is clearly observed. The optical conductivity
 thus allows to show that even in the metallic (deconfined) regime
 LL behavior is still present. Because of the wide frequency range
 accessible on which the power law is seen it also allows to extract the LL parameter $K_\rho$ reliably.
 (From \figcite{schwartz_electrodynamics} (Copyright 1998 by the American
Physical Society).)}
 \label{fig:opticala}
\end{figure}

\begin{figure}
 \centerline{\includegraphics[width=\normfig]{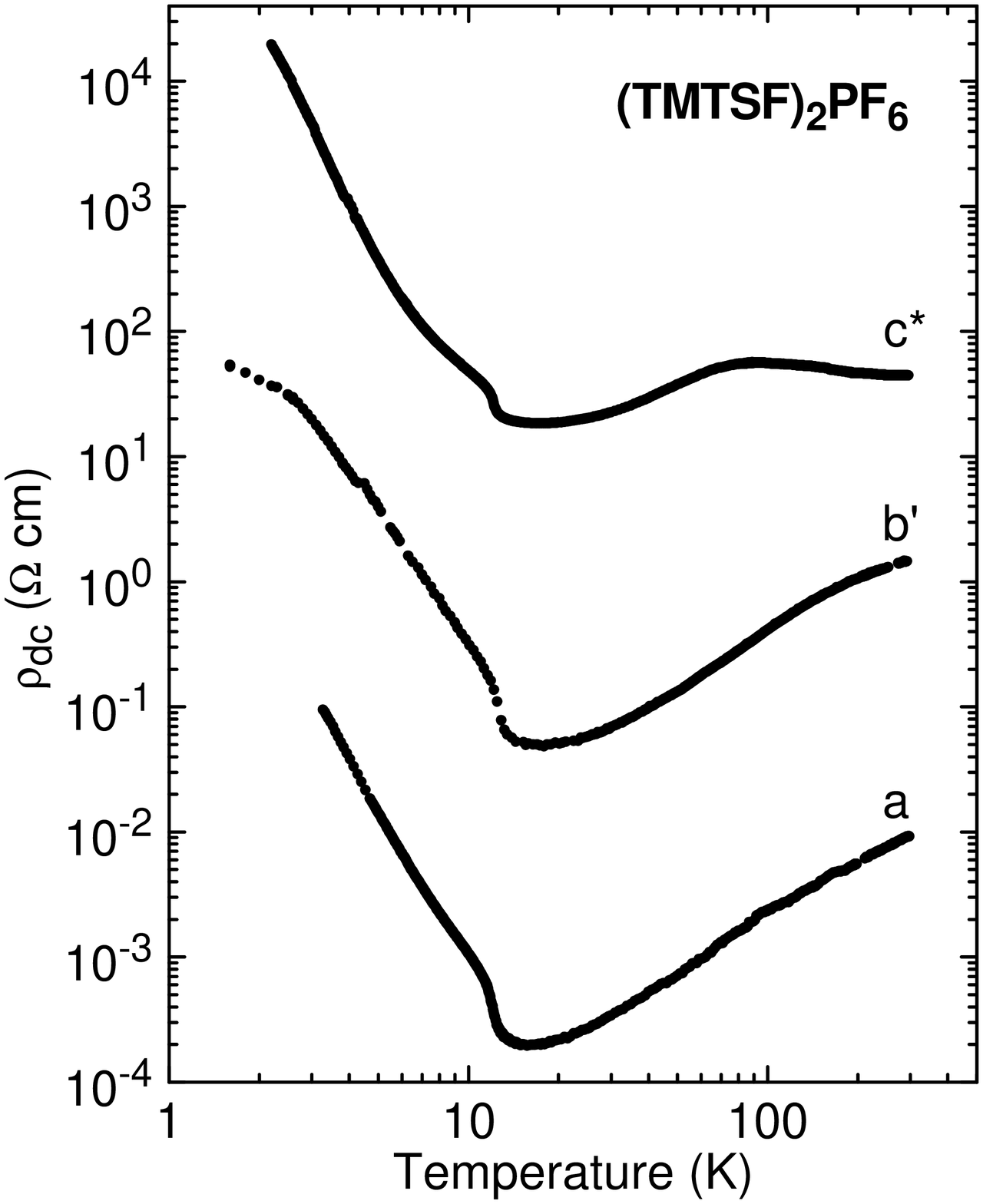}}
 \centerline{\includegraphics[width=\normfig]{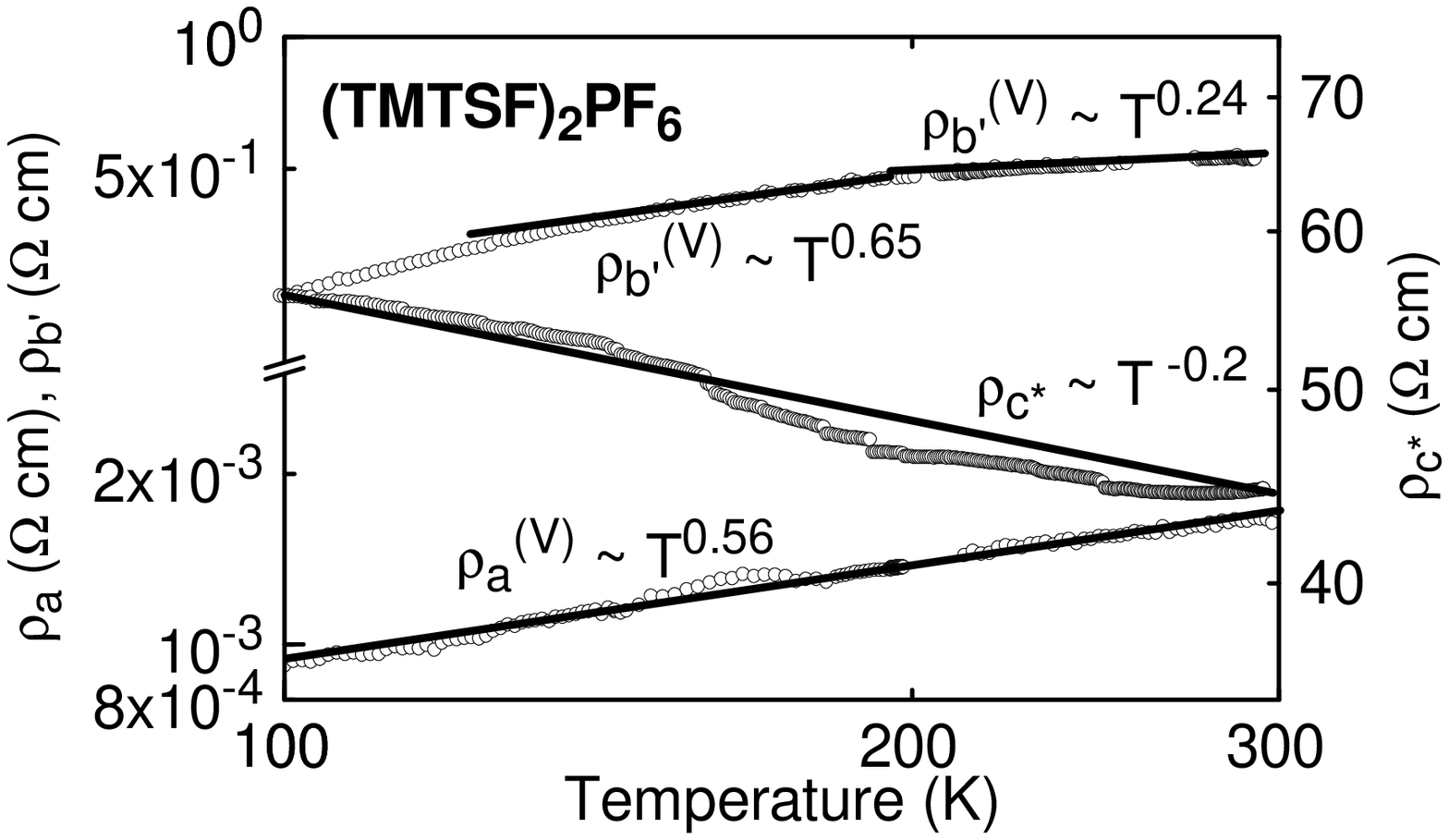}}
 \caption{(top) Conductivity of TMTSF$_2$PF$_6$ at constant pressure. The difference between the $a$ and $c$ axis at high
 temperature is directly visible and is a signature of the LL behavior. The crossover temperature
 scale is here slightly below $100K$ as indicated by the change of behavior of $\rho_c(T)$.
 (bottom) The conductivity, corrected to be the constant volume one is shown above the crossover scale $T^*$.
 Both the $\rho_a(T)$ and $\rho_c(T)$ temperature dependence are well compatible with the value of the LL parameter
 $K_\rho \sim 0.23$ extracted from the optics. The crossover in the $\rho_b(T)$ dependence is obviously much broader
 but the same qualitative tendency than in $\rho_c(T)$ can be seen. (From
 \figcite{dressel_transport_tmtsf} (Copyright 2005 by the American Physical Society).)}
 \label{fig:condpf6}
\end{figure}

Independent of any theory, a clear proof of the importance of
interactions for {\it both} the TMTTF and TMTSF compounds is
provided by the optical conductivity
\cite{dressel_optical_tmtsf,schwartz_electrodynamics}. The optical
conductivity shows a decreasing gap (of the order of 2000\ cm$^{-1}$
for the TMTTF$_2$(PF$_6$) to 200\ cm$^{-1}$ for TMTSF$_2$(PF$_6$).
Nearly (99\%) of the spectral weight is in this high-energy
structure. In the metallic compounds there is in addition a very
narrow Drude peak (see \cite{bourbonnais_book_organics} for
additional data). This clearly indicates that these compounds are
very far from simple Fermi liquids. Furthermore, the data of optical
conductivity can be compared with the theoretical calculations for a
one-dimensional Mott insulator (see \fref{sigacdel0}) as shown in
\fref{fig:opticala}.
The data above the gap fits very well the predicted power law LL
behavior (\ref{eq:accond}) above the gap and thus shows quite
convincingly that these compounds are indeed well described by a LL
theory at high energy \cite{schwartz_electrodynamics}. This was, to
the best of my knowledge, the first direct proof of a Luttinger
liquid behavior in an electronic system. This measurement also
allows to directly extract the Luttinger liquid parameter $K_\rho$.
A similar comparison can be done on the dc transport and gives also
good agreement
\cite{jerome_organic_review,jerome_review_chemrev,bourbonnais_book_organics}
with the predicted power law (\ref{eq:dccond}). Recent d.c. transport
data on (TMTSF)$_2$PF$_6$ \cite{dressel_transport_tmtsf} are also
shown in \fref{fig:condpf6}. From the data of \fref{fig:condpf6} one
sees that (TMTTF)$_2$PF$_6$ shows a quite consistent behavior with
the above theoretical description. For temperatures $T$ larger than
about $100K$ one recovers a power-law $\rho_a \propto T^{0.56}$
quite compatible, using (\ref{eq:dccond}) with the value of $K_\rho
= 0.22-0.23$ obtained from the optics.
Additional experiments both in
optics
\cite{vescoli_confinement_science,vescoli_photoemission_tmtsf,henderson_transverse_optics_organics,pashkin_pressure_optics_asf6}
and on dc transport
\cite{jerome_review_chemrev,bourbonnais_book_organics,dressel_transport_tmtsf}
have confirmed the LL nature, and thus the upper part of the phase
diagram of \fref{fig:deconf_phasediag}.

Quite remarkably, optical conductivity is in fact an excellent probe
of the LL behavior. Indeed if the control parameter is the
temperature several limitations are present. First the compounds are
quite compressible, and it is important to make pressure corrections
to compare the theoretical prediction (at constant volume) with the
experimental observations
\cite{jerome_organic_review,jerome_review_chemrev,bourbonnais_book_organics}.
Not taking into account this volume change with temperature was
initially responsible for the seemingly $T^2$ behavior observed till
room temperature. However the pressure correction is hard to perform
with great accuracy and more importantly the range of temperature
between $T^*$ and room temperature, which is in practice about the
maximum range available, is very limited (typically $100$--$300$K at
best in PF$_6$) making difficult a quantitative and convincing test
of a power law regime. Optics does not suffer from those limitations.
Since the parameter varied is the frequency, no such expansion
correction is needed. The range of energy that can be explored is
also much larger and limited only by the bandwidth of the system.
This allows for a fit of the power law on more than a decade.  It is
important to note that such an analysis of a.c. transport has also
been used with success in other types of one dimensional materials
\cite{lee_optics_chains_ybco}. It is thus a method of choice to
probe for the physics of these systems.

Indeed quite importantly the $a$-axis optical measurements described
above even allow for a quantitative determination
\cite{schwartz_electrodynamics} of the LL parameter $K_\rho$ and a
better understanding of the mechanism behind the Mott transition in
these materials. A fit of the frequency dependence of the
longitudinal conductivity (see \fref{fig:opticala}) can be performed
using (\ref{eq:accond}). A commensurability of order one ($n=1$)
does not allow for a consistent fit of both the exponent and the
Mott gap \cite{schwartz_electrodynamics}. Such a commensurability
would lead to $K_\rho \sim 1$, a nearly non-interacting system and
then to a gap much smaller than the one observed experimentally.
This indicates that, at least for the TMTSF members of the family,
the dominant umklapp comes from the quarter filled nature of the
band. Formula (\ref{eq:accond}) with $n=2$ thus yields $K_\rho
\simeq 0.23$, indicating quite strong electron--electron
interactions. Moreover this indicates that the finite range nature
of the interactions should be taken into account, with interactions
extending at least to nearest neighbors. A modelization of the
organics thus should not be done with a purely local Hubbard model,
with an interaction $U$, but also take into account at least the
nearest neighbor interaction $V$. The optical data is thus
consistent with an interpretation of the insulating state as a
quarter-filled Mott insulator, suggesting that contrarily to what
was commonly believed \cite{emery_umklapp_dimerization} the
dimerization plays little role at least in the TMTSF family. In the
TMTTF family, dimerization is larger and it is unclear there which
process is dominant. Note that because of the anions other
transitions can exist such as a ferroelectric transition
\cite{monceau_ferroelectric_tmttf,brazovskii_iscom_1d,brazovskii_book_organics}.
It is important to note that this also suggests
\cite{schwartz_electrodynamics,giamarchi_review_chemrev}, that a
very important effect of pressure, is not so much to affect the
dimerization, but to affect the hoppings and thus reduce the
interaction versus kinetic energy ratio. This makes the system
effectively less interacting with pressure. This prediction has been
recently confirmed on optical measurements under pressure where a
consistent decrease of the Luttinger parameter $K_\rho$ has been
observed with increasing pressure
\cite{pashkin_pressure_optics_asf6}. Given the fact that the value
of $K_\rho$ is very close to the critical value $K_\rho = 0.25$ for
which the quarter filled umklapp becomes irrelevant, such a
variation of $K_\rho$ can trigger a rapid variation of the Mott gap
upon application of pressure, or when going from the TMTTF members
to the TMTSF members. The importance of the quarter filled umklapp
in this family of compounds has been also clearly confirmed by
properties of parents compounds with a structure similar to the
Bechgaard salts
but that do not have dimerization
\cite{kato_quarter_synthesis,heuze_quarterfilled_refs,hiraki_quarter_measurements,%
itou_quarter_1d}. These compounds turned out to be Mott insulators
\cite{bourbonnais_book_organics}. Under pressure they share most of
the features of the Bechgaard and Fabre salts, indicating that the
same physics is at hand
\cite{heuze_quarterfilled_refs,itou_quarter_1d,itou_quarter_dimensional}.
It would be of course very interesting to further investigate the
phase diagram and the transport properties under pressure of these
compounds. In addition, since they share the same basis microscopic
features, it would be specially interesting to assert whether these
quarter filled systems also exhibit superconductivity under pressure
as in the Bechgaard salts.

Finally the last question that can be addressed by the transport
along the chains, is the one of the value of the crossover scales
$\Delta_\rho(t_\perp)$ or $T^*(t_\perp)$ as shown in
\fref{fig:deconf_phasediag}. $\Delta_\rho(t_\perp)$ is easily seen
from the upturn of the resistivity along $a$-axis or directly from
the optics for the insulating members of the family. In the metallic
regime $T^*(t_\perp)$ can be estimated by the crossover between a
$T^2$ behavior at low temperature to the nonuniversal power law
(\ref{eq:dccond}) corresponding to the LL at high temperatures. For
example this suggests a crossover scale of about $T^* \sim 100K$ for
(TMTSF)$_2$PF$_6$ as seen on \fref{fig:condpf6}. However a much more
precise determination of this scale is provided by a measure of the
transverse transport \cite{moser_conductivite_1d} that I now
examine.

\section{Coupled chains}
\label{sec:coupledchains}

Let us now investigate the effects that are direct consequences of
the coupling between the chains. There are of course the
deconfinement transition at $t_\perp^*$ and the two crossover scales
$\Delta(t_\perp)$ and $T^*(t_\perp)$. But the very fact that many
chains are presents means that transverse transport effects can be
probed as well, even in the high energy (LL) of the phase diagram of
\fref{fig:deconf_phasediag}. Such transverse transport is sensitive
on how electrons can tunnel from one of the chain to the other. It
thus reflects directly how well single particle excitations can
exist, and is therefore also a way to probe the LL nature of the
system. It is of course also a very sensitive way to address the
question of the dimensional crossover since one can expect a
drastically different type of transverse transport depending on
whether single particle excitations exists in the chain and thus can
hop or not.

Let me first discuss the LL region in \fref{fig:deconf_phasediag}.
In that region the hopping is incoherent between the chains. Thus
the transverse conductivity can be computed in the high temperature
or high frequency regime by an expansion in the perpendicular
hopping \cite{georges_organics_dinfiplusone}. One finds a power-law
either in frequency of temperature, controlled by the single
particle Green's function exponent. At finite temperatures
\begin{equation} \label{eq:taxis}
 \sigma_\perp (T\gg\omega) \propto T^{2\alpha-1}
\end{equation}
for $k_B T\gg E_{\rm cross}$, while at high-frequency
($\hbar\omega\gg E_{\rm cross}$), one gets
\begin{equation} \label{eq:caxis}
 \sigma_\perp(\omega\gg T)\propto \omega^{2\alpha-1}
\end{equation}
where $\alpha = \zeta - 1 = \frac14(K_\rho+K_\rho^{-1}) - \frac12$
is the exponent in the single particle density of states. $E_{\rm
cross}$ is the scale at which this expansion breaks down, either
$\Delta_\rho(t_\perp)$ or $T^*(t_\perp)$ as given by the dotted and
dashed lines in \fref{fig:deconf_phasediag}. Note that in the regime
where chains are in the LL state if one takes $K_\rho \sim 0.23$ as
given by the measurement of the intra chain transport, the
transverse conductivity \emph{decreases} with decreasing temperature
or frequency. One has thus a very different behavior of the intra
and interchain transport. This is to be contrasted from a normal
Fermi liquid regime, where one can expect similar temperature
dependencies in both directions. The change of behavior can thus be
used to detect the dimensional crossover scale $T^*$. This is quite
clear on \fref{fig:condpf6}.

Observed optical conductivity along $c$-axis
\cite{henderson_transverse_optics_organics} is compatible with the
power law growth of (\ref{eq:caxis}) and a value of $K_\rho \sim
0.23$, as determined by the in-chain transport. But clearly much
more experiments would be needed since the measurement is extremely
difficult and the data not at the same level of accuracy than the
intrachain transport. At the price of the pressure correction the dc
transport can also be used. For (TMTTF)$_2$PF$_6$, as shown in
\fref{fig:condpf6}, the temperature dependence of the in-chain
conductivity gives, with (\ref{eq:dccond}), a value of $K_\rho=
0.22-0.23$ well compatible with the one from the optics. In the same
way $\rho_{c}\propto T^{-0.2}$ and (\ref{eq:taxis}) gives $\alpha =
0.69$ again well compatible with $K_{\rho} = 0.22$. Note that the
fact that the dc transport is at least qualitatively reproduced with
the \emph{same} value of $K_\rho$ gives strong credence to a LL
physics interpretation of the data. The interpretation of $\rho_{b}$
is more complex since no simple power-law is seen over the entire
temperature range. However a change in the anisotropy behavior takes
place above $T^* \sim 100K$ in \fref{fig:condpf6} and the fit to a
power-law shows a clear downturn of the exponent. A possible
interpretation of the data is thus as being in a crossover regime
between the low-temperature Fermi liquid one and the
high-temperature Luttinger liquid. Note that it is reasonable to
expect a much larger crossover region for the $b^{\prime}$-axis
transport, than for the $c^*$ axis given the much higher value of
the transfer integral in this direction. Despite this general
agreement, a quantitative understanding of the transport along $b$
clearly requires further work both for transport
\cite{dressel_transport_tmtsf} and for optics
\cite{henderson_transverse_optics_organics}.

The temperature dependence below $T \approx 100$~K is the same for
the $a$ and $b^{\prime}$ directions implying a similar transport
mechanism, and the anisotropy ratio corresponds roughly to the
expected band structure value. Below this scale, the dc resistivity
follows a power-law $\rho_a, \rho_{b} \propto T^2$, as expected for
a Fermi liquid. Note that although the system is now two dimensional
because the hopping in the $b$ direction is now coherent, the
temperature is still much larger than the hopping in the $c$
direction $T \gg t_c$. Thus one can still use formula
(\ref{eq:taxis}) but putting $\alpha = 0$ as would befit a Fermi
liquid. This gives $\rho_{c} \propto T$ which is effectively
consistent with the experimental data. The experiments in
(TMTSF)$_2$PF$_6$ thus directly confirms the above theoretical
analysis and the Luttinger liquid behavior. The crossover scale can
be experimentally determined to be $T^* \sim 100K$ for the case of
PF$_6$. For the case of TMTSF$_2$ClO$_4$ a similar analysis suggests
a much higher value $T^* \geq 200K$ \cite{dressel_transport_tmtsf}.
However several problems remain with this compound, in particular
concerning the high temperature resistance anisotropy which is much
smaller than normally expected. More experimental data is clearly
needed in that case. The crossover scale can also be followed under
pressure (see Fig.~1.8 of \cite{bourbonnais_book_organics}). Another
way to determine the crossover scale is provided by the optical $b$
axis conductivity since coherent hopping between the chains
manifests itself as the appearance of a Drude peak in the $b$ axis
conductivity \cite{henderson_transverse_optics_organics}.

Another measurement that can in principle probe the nature of the
Luttinger liquid is of course the Hall effect. Indeed in a Fermi
liquid the Hall effect is essentially a measure of Fermi surface
properties. At low temperatures (in the HDM part of the diagram) the
Hall effect can be quite successfully described in this framework
\cite{yakovenko_phenomenological_model_th,yakovenko_phenomenological_model_exp,yakovenko_book_organics}.
At temperatures larger that $T^*$ one could expect the Hall effect
to reflect again the nature of the interactions in the LL phase.
However both the theory and the experiments concerning this quantity
are more complicated. Experiments with the magnetic field along the
$c$ axis \cite{moser_hall_1d,Korin-Hamzic_hall_reo4} observe a weak
temperature dependence while field along $a$ leads to an essentially
temperature independent Hall effect \cite{mihaly_hall_1d}. On the
theoretical side, quite surprisingly it was shown that in the
absence of scattering along the chains the Hall effect in a
Luttinger liquid does not show any trace of the interactions and is
equal to the band value
\cite{lopatin_hall_luttinger,lopatin_q1d_magnetooptical} $R_h^0$.
Including the scattering along the chains was done recently
\cite{leon_hall_quasi1d_conf,leon_hall_quasi1d_short} for the case
of the half-filled umklapp and magnetic field perpendicular to the
chains, leading to a Hall effect behaving as
\begin{equation}\label{RH_result}
        R_h=R_h^0\left[1-A\left(\frac{g_{\frac12}}{\pi v_{\text{F}}}\right)^2
        \left(\frac{T}{W}\right)^{3K_{\rho}-3}\right],
\end{equation}
where $A$ is a dimensionless constant and $W$ the bandwidth of the
material. This shows that some temperature dependence is to be
expected in the Luttinger regime, in qualitative agreement with the
observations \cite{moser_hall_1d,Korin-Hamzic_hall_reo4}.

\begin{figure}
 \centerline{\includegraphics[width=\datafig]{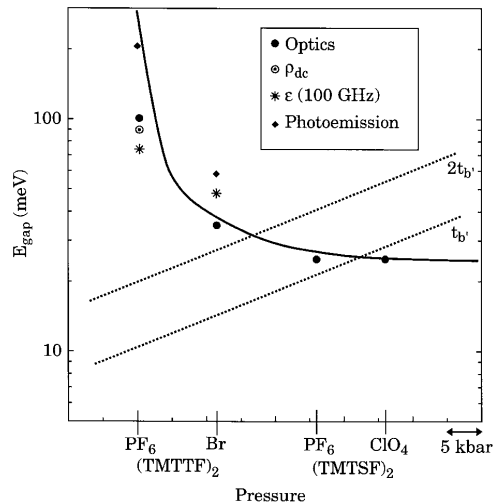}}
 \caption{A comparison of the measured gap in the optical
 conductivity with the interchain hopping. The change of behavior
 from insulator to metallic occurs when the two quantities are of
 the same order of magnitude showing that the difference between
 the various members of the TM families is indeed linked to a
 deconfinement transition.  (From
 \figcite{vescoli_photoemission_tmtsf} (Copyright 2000 by EDP Sciences).)}
 \label{fig:confine}
\end{figure}

As one can obtain the crossover temperature $T^*$, one can also
determine the deconfinement critical value $t_\perp^*$ from the
transport measurements. This can be done for example by monitoring
the occurrence of metallic behavior in the $a$-axis (see Fig.~1.8 of
\cite{bourbonnais_book_organics}). In addition, a measure of the gap
extracted from the optical conductivity shows that the change of
nature between insulating to metallic behavior occurs when the
observed gap is roughly of the order of magnitude of the interchain
hopping \cite{vescoli_confinement_science} (see \fref{fig:confine}).
On the theoretical side understanding quantitatively the
deconfinement transition and even the crossover scale $T^*$ for
deconfined systems is a major theoretical challenge. In the absence
of commensurability, the crossover scale between a LL and the HDM
can be determined by looking at the renormalization of the
interchain hopping
\cite{brazovskii_transhop,bourbonnais_transhop,wen_coupled_chains,%
bourbonnais_couplage,yakovenko_manychains,boies_couplage,schulz_moriond}.
If one neglects the renormalization of $\alpha$ by the interchain
hopping, then one has \cite{bourbonnais_transhop}
\begin{equation} \label{eq:singlepartcross}
 T^* \sim W \left(\frac{t_\perp}{W}\right)^{\frac{1}{1 - \alpha}}
\end{equation}
For the non-interacting case $\alpha = 0$ and one recovers $E^* \sim
t_\perp$. Since $\alpha$ for an interacting system is always
positive, the scale at which the dimensional crossover takes place
is always \emph{smaller} than for free fermions. Interactions thus
tend to make the system more one-dimensional. This reduction of the
crossover scale comes again from the fact that in a Luttinger liquid
{\it single particle excitations} are strongly suppressed. For the
commensurate case, the scale $T^*$ must be computed in presence of
the umklapp term, and is thus dependent also on the deconfinement
scale. Unfortunately the RG study, although it provides the scale at
which the LL is unstable cannot carry easily through the low
temperature HDM. What is the nature of this phase is thus still a
major challenge. Even if it is a Fermi liquid, since this Fermi
liquid stems from the high temperature non-Fermi liquid phase, its
features are certainly quite special. In particular the
quasiparticle residue $Z$ and lifetime of the quasiparticles could
in principle retain the memory of the strong correlations that
existed in the one dimensional phase
\cite{georges_organics_dinfiplusone,giamarchi_review_chemrev}. In
addition since the strength of the hopping depends on the transverse
momentum $k_\perp$ these quantities could be varying on the Fermi
surface and lead to the presence of hot spots
\cite{zheleznyak_hot_spots}.

Besides the RG analysis various methods have been tried to tackle the
deconfinement transition. This is a difficult problem and much less
is known than for the dimensional crossover. Both scaling arguments
\cite{giamarchi_mott_shortrev} and study of two chain systems
\cite{suzumura_confinement_ladder,tsuchiizu_confinement_ladder,tsuchiizu_confinement_ladder_long,%
tsuchiizu_confinement_spinful,lehur_ladder_crossover}, showed the
importance of the energy scale $T^*$ in comparison with the Mott gap
$\Delta_\rho^0$ in the absence of $t_\perp$. A rule of thumb to get
the position of this deconfinement transition is to compare the two
scales $T^*$ and $\Delta_\rho^0$. Thus, roughly if $T^*
> \Delta_\rho^0$ one is deconfined, whereas for $T^* < \Delta_\rho^0$ the gap
wins and the chains are confined, only allowing for two particles
hopping. Of course, this is only a rule of thumb and one should, in
principle, solve the full coupled problem to obtain the critical
value $t_\perp^*$ at which deconfinement occurs. No full solution of
this problem exists so far. An RG analysis properly incorporating
the umklapp terms and the interchain coupling up to the
deconfinement transition is quite difficult to realize in a
controlled way since both phase correspond to strong coupling fixed
points. An RPA treatment \cite{essler_rpa_quasi1d} of the hopping
does produce an insulator-metal transition via the formation of
pockets on the Fermi surface. It however neglects any feedback of
the hopping on the one dimensional gap itself and thus grossly
overestimate the position of the transition. It also cannot give a
full deconfinement with an open Fermi surface, since the
one-dimensional gap never closes. A quite promising method is a mean
field approach (ch-DMFT) treating the chains as an effective bath
\cite{arrigoni_tperp_resummation,georges_organics_dinfiplusone,arrigoni_tperp_resummation_prb,%
biermann_dmft1d_hubbard_short,biermann_oned_crossover_review,berthod_deconfinement_spinless}.
This method shows clearly the deconfinement transition and gives
access to some of the properties of the HDM phase beyond the
transition. An analysis has been performed for the half filled case
and I refer the reader to
\cite{biermann_dmft1d_hubbard_short,biermann_oned_crossover_review}
for more details. The full analysis of the quarter filled band,
relevant for the quasi-one dimensional organics still remains to be
done. A caricature of this case, corresponding roughly to a very
large on-site interaction $U$ and a moderate nearest neighbor
interaction $V$ can however be performed by considering an half
filled band of \emph{spinless} fermions
\cite{berthod_deconfinement_spinless}. In that case a strong
depletion of the Mott gap with increasing $t_\perp$ has been
observed and the deconfinement transition has been analyzed. In such
an approach deconfinement occurs first through formations of pockets
at a first critical value $t_{c1}$ but then at a slightly larger
value $t_{c2}$ a full open Fermi surface is recovered. In the HDM
phase, effects of the interactions can still be felt, as in
particular the presence of hot spots on the Fermi surface
\cite{berthod_deconfinement_spinless}. Despite these progress more
work both theoretical and experimental is needed to completely
understand the deconfinement transition. On the experimental side,
it would of course be particularly interesting to have information
on single particle excitations. Unfortunately photoemission or STM
tunnelling experiments on the organic conductors seems to be
difficult due to the ionic nature of the systems and the surface
problems it entails. Some results consistent with Mott physics and
Luttinger liquids were observed, in particular the Mott gaps
\cite{dardel_photoemission,zwick_bechgaard_arpes,vescoli_photoemission_tmtsf}.
But given the very large energy scales at which for example a
depletion of the density of state has been observed, interpretation
of these results in terms of LL should be taken with a grain of
salt.

The physics below the dimensional crossover scale also remains to be
fully understood. However we at least now know reliably from the
above mentioned transport experiments the crossover scale to the
HDM. This allows to sort out effects that can be attributed to the
one-dimensional behavior from those that one has to understand in a
more conventional high dimensional system. For the TMTSF members the
crossover scale being at least of the order of $\sim 100 K$ (for
PF$_6$) one can expect that a Fermi liquid approach should be a
useful starting point much below this temperature. Indeed, Fermi
liquid theory has been quite successful in explaining many of the
\emph{low temperature} ordered phase and properties of these
compounds (see e.g the chapters on the field induced spin density
waves in this book). On the other hand some correlation effects
beyond simple Fermi liquids still persists even way below $T^*$.
This is clear both from the above mentioned theoretical calculations
and, from the experimental side, by the existence of anomalies such
as the ones in NMR \cite{bourbonnais_rmn,yu_nmr_organics}. Such
anomalies, being below $T^*$ cannot be attributed to Luttinger
liquid behavior. What is their explanation is still an open and a
challenging issue.

\section{Conclusions and perspectives}
\label{sec:conclusions}

I have presented in this review the main concepts and questions
relevant to tackle the normal phase physics of quasi-one dimensional
systems. The most important ones for isolated chains are the
Luttinger liquid theory and the Mott insulating physics which is
quite special in one dimension. For quasi-one dimensional systems an
extremely rich physics stems from the coupling between the chains.
Its most spectacular expression is the presence of a deconfinement
transition between a one dimensional insulator and a high
dimensional metal. The quasi-one dimensional organic conductors,
provide wonderful systems to investigate these phenomena. But this
question is pertinent for many experimental systems and is in
particular now investigated in systems such as cold atomic gases as
well
\cite{ho_deconfinement_coldatoms,stoferle_tonks_optical,cazalilla_coupled_fermions}.

I have shown here how a good theoretical understanding of the
transport properties allows to probe the unconventional physics of
these systems. In particular the transport has allowed to prove the
Luttinger liquid properties of the quasi-one dimensional organics.
It was also instrumental in determining the crossover scale (about
$100K$ for PF$_6$) between the one-dimensional LL properties and the
more conventional high dimensional metallic ones. It also forced to
a reexamination of the mechanism underlying the insulating behavior
in this compounds showing clearly that the quarter filled nature of
the system is enough in itself to lead to an insulating behavior.
These findings were confirmed by the existence of non-dimerized
compounds with Mott insulating properties. Thanks to these recent
progress we have now a consistent description of the relevant
properties and energies scales in the quasi-one dimensional
organics. This framework now provides a solid reference to focuss on
the important still unsolved questions.

Of course the remaining challenges are numerous. The low energy
phase properties are still largely not understood, and if some are
strongly reminiscent of the ones of a Fermi liquids, some deviate
markedly from them, such as the NMR. Even if one has a Fermi liquid
it is unlikely to be a plain vanilla one, since it will remember
that it stemmed from a low dimensional highly interacting system.
This can manifests itself, as is apparent on e.g. the mean field
solution, by the variation of the Fermi liquid parameters along the
Fermi surface. Much work thus remains to be done to understand this
phase, and possibly have a clue on the consequences on the ordered
phases, such as the superconducting one. In a similar way the field
opened by the new non-dimerized compound must be explored. In
particular it is important to determine whether they have indeed the
same properties than the TMTTF and TMTSF salts under suitable
pressure. No doubts that transport both a.c. and d.c. will prove a
useful tool to tackle these systems too. Finally of course, if any
for the theorist, since for now no chemist has managed to dope such
systems, it would be crucial to complete the phase diagram by adding
the doping axis, since most of the questions raised here become even
more crucial for doped Mott insulators.

\section{Acknowledgements}

Many people have contributed, directly via enjoyable collaborations,
or indirectly via scientific discussions to the work presented here
or more generally to my own understanding of the field. The list of
the persons I would like to thank would be too long to be given
here, but I would like to specially mention: L. Degiorgi, M.
Dressel, A. Georges, D. J\'erome, H.J. Schulz. Part of this work was
supported by the Swiss national fund for research under MaNEP and
division II.


\begin{thebibliography}{10}

\bibitem{landau_fermiliquid_theory_static}
L.~D. Landau, Sov. Phys. JETP {\bf 3},  920  (1957).

\bibitem{nozieres_book}
P. Nozieres, {\em Theory of Interacting Fermi Systems} (Benjamin,
New York,
  1961).

\bibitem{abrikosov_book}
A.~A. Abrikosov, L.~P. Gorkov, and I.~E. Dzyaloshinski, {\em Methods
of Quantum
  Field Theory in Statistical Physics} (Dover, New York, 1963).

\bibitem{giamarchi_book_1d}
T. Giamarchi, {\em Quantum Physics in One Dimension} (Oxford
University Press,
  Oxford, 2004).

\bibitem{giamarchi_review_chemrev}
T. Giamarchi, Chem. Rev. {\bf 104},  5037  (2004).

\bibitem{bourbonnais_book_organics}
C. Bourbonnais and D. Jerome, 2006, in this book.

\bibitem{giamarchi_umklapp_1d}
T. Giamarchi, Phys. Rev. B {\bf 44},  2905  (1991).

\bibitem{giamarchi_mott_shortrev}
T. Giamarchi, Physica B {\bf 230-232},  975  (1997).

\bibitem{dressel_optical_tmtsf}
M. Dressel, A. Schwartz, G. Gr{\"u}ner, and L. Degiorgi, Phys. Rev.
Lett. {\bf
  77},  398  (1996).

\bibitem{schwartz_electrodynamics}
A. Schwartz {\it et~al.}, Phys. Rev. B {\bf 58},  1261  (1998).

\bibitem{emery_umklapp_dimerization}
V.~J. Emery, R. Bruinsma, and S. Barisic, Phys. Rev. Lett. {\bf 48},
1039
  (1982).

\bibitem{kato_quarter_synthesis}
R. Kato, H. Kobayashi, and A. Kobayashi, J. Am. Chem. Soc. {\bf
111},  5224
  (1989).

\bibitem{heuze_quarterfilled_refs}
K. Heuz{\'e} {\it et~al.}, Adv. Mater. {\bf 15},  1251  (2003).

\bibitem{biermann_dmft1d_hubbard_short}
S. Biermann, A. Georges, A. Lichtenstein, and T. Giamarchi, Phys.
Rev. Lett.
  {\bf 87},  276405  (2001).

\bibitem{gorkov_sdw_tmtsf}
L.~P. Gorkov, Physica B {\bf 230-232},  970  (1997).

\bibitem{bourbonnais_rmn}
C. Bourbonnais {\it et~al.}, J. Phys. (Paris) Lett. {\bf 45},  L755
(1984).

\bibitem{behnia_tmtsfclo4_rmn}
K. Behnia {\it et~al.}, Phys. Rev. Lett. {\bf 74},  5272  (1995).

\bibitem{luther_chaine_xxz}
A. Luther and I. Peschel, Phys. Rev. B {\bf 12},  3908  (1975).

\bibitem{schulz_hubbard_exact}
H.~J. Schulz, Phys. Rev. Lett. {\bf 64},  2831  (1990).

\bibitem{ogata_tj}
M. Ogata, M.~U. Luchini, S. Sorella, and F.~F. Assaad, Phys. Rev.
Lett. {\bf
  66},  2388  (1991).

\bibitem{mila_hubbard_etendu}
F. Mila and X. Zotos, Europhys. Lett. {\bf 24},  133  (1993).

\bibitem{kawakami_hubbard}
N. Kawakami and S.~K. Yang, Phys. Lett. A {\bf 148},  359  (1990).

\bibitem{kawakami_bethe_U<0}
N. Kawakami and S.~K. Yang, Phys. Rev. B {\bf 44},  7844  (1991).

\bibitem{haldane_bosonisation_spin}
F.~D.~M. Haldane, J. Phys. C {\bf 12},  4791  (1979).

\bibitem{haldane_bosonisation}
F.~D.~M. Haldane, J. Phys. C {\bf 14},  2585  (1981).

\bibitem{Haldane_Luttinger}
F.~D.~M. Haldane, Phys. Rev. Lett. {\bf 45},  1358  (1980).

\bibitem{lieb_hubbard_exact}
E.~H. Lieb and F.~Y. Wu, Phys. Rev. Lett. {\bf 20},  1445  (1968).

\bibitem{dzyaloshinskii_umklapp}
I.~E. Dzyaloshinskii and A.~I. Larkin, Sov. Phys. JETP {\bf 34},
422  (1972).

\bibitem{ziman_phonon_book}
J.~M. Ziman, {\em Electrons and Phonons} (Clarendon, Oxford, 1962).

\bibitem{giamarchi_curvature}
T. Giamarchi and A.~J. Millis, Phys. Rev. B {\bf 46},  9325  (1992).

\bibitem{schulz_mott_revue}
H.~J. Schulz,  in {\em Strongly Correlated Electronic Materials: The
Los Alamos
  Symposium 1993}, edited by K.~S. {Bedell {\it et al.}} (Addison--Wesley,
  Reading, MA, 1994), p.\ 187.

\bibitem{jerome_review_chemrev}
D. J{\'e}rome, Chem. Rev. {\bf 104},  5565  (2004).

\bibitem{schulz_conductivite_1d}
H.~J. Schulz, Phys. Rev. Lett. {\bf 64},  2831  (1990).

\bibitem{giamarchi_attract_1d}
T. Giamarchi, Phys. Rev. B {\bf 46},  342  (1992).

\bibitem{controzzi_cond1d_form}
D. Controzzi, F.~H.~L. Essler, and A.~M. Tsvelik, Phys. Rev. Lett.
{\bf 86},
  680  (2001).

\bibitem{jeckelmann_dmrg_conductivity_hubbard}
E. Jeckelmann, F. Gebhard, and F.~H.~L. Essler, Phys. Rev. Lett.
{\bf 85},
  3910  (2000).

\bibitem{rosch_conservation_1d}
A. Rosch and N. Andrei, Phys. Rev. Lett. {\bf 85},  1092  (2000).

\bibitem{jerome_organic_review}
D. J{\'e}rome,  in {\em Organic Superconductors: From
{(TMTSF)$_2$PF$_6$} to
  Fullerenes} (Marcel Dekker, New York, 1994), Chap.~10, p.\ 405.

\bibitem{dressel_transport_tmtsf}
M. Dressel {\it et~al.}, Phys. Rev. B {\bf 71},  075104  (2005).

\bibitem{vescoli_confinement_science}
V. Vescoli {\it et~al.}, Science {\bf 281},  1191  (1998).

\bibitem{vescoli_photoemission_tmtsf}
V. Vescoli {\it et~al.}, Eur. Phys. J. B {\bf 13},  503  (2000).

\bibitem{henderson_transverse_optics_organics}
W. Henderson {\it et~al.}, Eur. Phys. J. B {\bf 11},  365  (1999).

\bibitem{pashkin_pressure_optics_asf6}
D. Pashkin, M. Dressel, and C.~A. Kuntscher, 2006, cond-mat/0608242.

\bibitem{lee_optics_chains_ybco}
Y.-S. Lee, K. Segawa, Y. Ando, and D.~N. Basov, Phys. Rev. Lett.
{\bf 94},
  137004  (2005).

\bibitem{monceau_ferroelectric_tmttf}
P. Monceau, F. Nad, and S. Brazovskii, Phys. Rev. Lett. {\bf 86},
4080
  (2001).

\bibitem{brazovskii_iscom_1d}
S. Brazovskii, J. Phys. IV {\bf 114},  9  (2004).

\bibitem{brazovskii_book_organics}
S. Brazovskii, 2006, in this book.

\bibitem{hiraki_quarter_measurements}
K. Hiraki and K. Kanoda, Phys. Rev. B {\bf 54},  R17276  (1996).

\bibitem{itou_quarter_1d}
T. Itou {\it et~al.}, Phys. Rev. Lett. {\bf 93},  216408  (2004).

\bibitem{itou_quarter_dimensional}
T. Itou {\it et~al.}, Phys. Rev. B {\bf 72},  113109  (2005).

\bibitem{moser_conductivite_1d}
J. Moser {\it et~al.}, Eur. Phys. J. B {\bf 1},  39  (1998).

\bibitem{georges_organics_dinfiplusone}
A. Georges, T. Giamarchi, and N. Sandler, Phys. Rev. B {\bf 61},
16393
  (2000).

\bibitem{yakovenko_phenomenological_model_th}
V.~M. Yakovenko, Synth. Metal {\bf 103},  2202  (1999).

\bibitem{yakovenko_phenomenological_model_exp}
V.~M. Yakovenko and A.~T. Zheleznyak, Synth. Metal {\bf 120},  1083
(2001).

\bibitem{yakovenko_book_organics}
V.~M. Yakovenko, 2006, in this book.

\bibitem{moser_hall_1d}
J. Moser {\it et~al.}, Phys. Rev. Lett. {\bf 84},  2674  (2000).

\bibitem{Korin-Hamzic_hall_reo4}
B. Korin-Hamzi\'c {\it et~al.}, Phys. Rev. B {\bf 67},  014513
(2003).

\bibitem{mihaly_hall_1d}
G. Mihaly, I. Kezsmarsky, F. Zamborsky, and L. Forro, Phys. Rev.
Lett. {\bf
  84},  2670  (2000).

\bibitem{lopatin_hall_luttinger}
A. Lopatin, Phys. Rev. B {\bf 57},  6342  (1997).

\bibitem{lopatin_q1d_magnetooptical}
A. Lopatin, A. Georges, and T. Giamarchi, Phys. Rev. B {\bf 63},
075109
  (2001).

\bibitem{leon_hall_quasi1d_conf}
G. Leon and T. Giamarchi, J. Low Temp. Phys {\bf 142},  315  (2006).

\bibitem{leon_hall_quasi1d_short}
G. Leon, C. Berthod, and T. Giamarchi, 2006, cond-mat/0608427.

\bibitem{brazovskii_transhop}
S. Brazovskii and V. Yakovenko, J. Phys. (Paris) Lett. {\bf 46},
L111  (1985).

\bibitem{bourbonnais_transhop}
C. Bourbonnais and L.~G. Caron, Physica {\bf 143B},  450  (1986).

\bibitem{wen_coupled_chains}
X.~G. Wen, Phys. Rev. B {\bf 42},  6623  (1990).

\bibitem{bourbonnais_couplage}
C. Bourbonnais and L.~G. Caron, Int. J. Mod. Phys. B {\bf 5},  1033
(1991).

\bibitem{yakovenko_manychains}
V.~M. Yakovenko, JETP Lett. {\bf 56},  510  (1992).

\bibitem{boies_couplage}
D. Boies, C. Bourbonnais, and A.-M.~S. Tremblay, Phys. Rev. Lett.
{\bf 74},
  968  (1995).

\bibitem{schulz_moriond}
H.~J. Schulz,  in {\em Correlated Fermions and Transport in
Mesoscopic
  Systems}, edited by T. Martin, G. Montambaux, and J. {Tran Thanh Van}
  (Editions fronti\`eres, Gif sur Yvette, France, 1996), p.\ 81.

\bibitem{zheleznyak_hot_spots}
A.~T. Zheleznyak and V.~M. Yakovenko, Synth. Metal {\bf 70},  1005
(1995).

\bibitem{suzumura_confinement_ladder}
Y. Suzumura, M. Tsuchiizu, and G. Gr{\"u}ner, Phys. Rev. B {\bf 57},
R15040
  (1998).

\bibitem{tsuchiizu_confinement_ladder}
M. Tsuchiizu, Y. Suzumura, and T. Giamarchi, Prog. Theor. Phys. {\bf
101},  763
   (1999).

\bibitem{tsuchiizu_confinement_ladder_long}
M. Tsuchiizu and Y. Suzumura, Phys. Rev. B {\bf 59},  12326  (1999).

\bibitem{tsuchiizu_confinement_spinful}
M. Tsuchiizu, P. Donohue, Y. Suzumura, and T. Giamarchi, Eur. Phys.
J. B {\bf
  19},  185  (2001).

\bibitem{lehur_ladder_crossover}
K. {Le Hur}, Phys. Rev. B {\bf 63},  165110  (2001).

\bibitem{essler_rpa_quasi1d}
F.~H.~L. Essler and A.~M. Tsvelik, Phys. Rev. B {\bf 65},  115117
(2002).

\bibitem{arrigoni_tperp_resummation}
E. Arrigoni, Phys. Rev. Lett. {\bf 83},  128  (1999).

\bibitem{arrigoni_tperp_resummation_prb}
E. Arrigoni, Phys. Rev. B {\bf 61},  7909  (2000).

\bibitem{biermann_oned_crossover_review}
S. Biermann, A. Georges, T. Giamarchi, and A. Lichtenstein,  in {\em
Strongly
  Correlated Fermions and Bosons in Low Dimensional Disordered Systems}, edited
  by I.~V. {Lerner {\it et al.}} (Kluwer Academic Publishers, Dordrecht, 2002),
  p.\ 81, cond-mat/0201542.

\bibitem{berthod_deconfinement_spinless}
C. Berthod, T. Giamarchi, S. Biermann, and A. Georges, 2006,
cond-mat/0602304.

\bibitem{dardel_photoemission}
B. Dardel {\it et~al.}, Europhys. Lett. {\bf 24},  687  (1993).

\bibitem{zwick_bechgaard_arpes}
F. Zwick {\it et~al.}, Phys. Rev. Lett. {\bf 79},  3982  (1997).

\bibitem{yu_nmr_organics}
W. Yu {\it et~al.}, Int. Review of Modern Physics B {\bf 16},  3090
(2002).

\bibitem{ho_deconfinement_coldatoms}
A.~F. Ho, M.~A. Cazalilla, and T. Giamarchi, Phys. Rev. Lett. {\bf
92},  130405
   (2004).

\bibitem{stoferle_tonks_optical}
T. St{\"o}ferle {\it et~al.}, Phys. Rev. Lett. {\bf 92},  130403
(2004).

\bibitem{cazalilla_coupled_fermions}
M.~A. Cazalilla, A.~F. Ho, and T. Giamarchi, Phys. Rev. Lett. {\bf
95},  226402
   (2005).

\end{thebibliography}
%

\printindex
\end{document}